\documentclass[useAMS,usenatbib]{mn2e}
\usepackage{epsfig}
\usepackage{deluxetable} 

\title[Multifrequency studies of the NLSy1 SBS\,0846$+$513]{Multifrequency studies of the narrow-line Seyfert 1 galaxy SBS\,0846$+$513} \author[F. D'Ammando, M. Orienti, J. Finke, et
al.]{F. D'Ammando$^{1,2,3}$\thanks{E-mail: dammando@ira.inaf.it}, M. Orienti$^{1,4}$, J. Finke$^{5}$, C. M. Raiteri$^{6}$, E. Angelakis$^{7}$, \newauthor L. Fuhrmann$^{7}$, M. Giroletti$^{1}$, T. Hovatta$^{8}$, V. Karamanavis$^{7}$, W. Max-Moerbeck$^{8}$, \newauthor I. Myserlis$^{7}$, A. C. S. Readhead$^{8}$, J. L. Richards$^{9}$\\
$^{1}$INAF - Istituto di Radioastronomia, Via Gobetti 101, I-40129 Bologna, Italy\\
$^{2}$Dipartimento di Fisica, Universit\`a degli Studi di Perugia, Via A. Pascoli, I-06123 Perugia, Italy \\
$^{3}$INFN Sezione di Perugia, Via A. Pascoli, I-06123 Perugia, Italy \\
$^{4}$Dipartimento di Astronomia, Universit\`a di Bologna, Via Ranzani 1, I-40127 Bologna, Italy \\ 
$^{5}$U.S. Naval Research Laboratory, Code 7653, 4555 Overlook Ave. SW, Washington, DC 20375-5352, USA \\
$^{6}$INAF - Osservatorio Astrofisico di Torino, Via Osservatorio 20, I-10025 Pino Torinese (TO), Italy \\
$^{7}$Max-Planck-Institute f\"ur Radioastronomie, Auf dem H\"ugel 69, D-53121 Bonn, Germany \\
$^{8}$Cahill Center for Astronomy and Astrophysics, California Institute of Technology, 1200 E. California Blvd, Pasadena, CA 91125, USA \\
$^{9}$Department of Physics, Purdue University, 525 Northwestern Avenue, West Lafayette, IN 47907, USA
}
\begin{document}

\date{Accepted. Received; in original form}

\maketitle

\label{firstpage}

\begin{abstract}
The narrow-line Seyfert 1 galaxy SBS\,0846$+$513 was first detected by the
Large Area Telescope (LAT) on-board {\em Fermi} in 2011 June--July when it
underwent a period of flaring activity. Since then, as {\em Fermi} continues to accumulate data on
this source, its flux has been monitored on a daily basis. Two further $\gamma$-ray
flaring episodes from SBS\,0846$+$513 were observed in 2012 May and August,
reaching a daily peak flux integrated above 100 MeV of
(50$\pm$12)$\times$10$^{-8}$ ph cm$^{-2}$ s$^{-1}$, and (73$\pm$14)$\times$10$^{-8}$ ph cm$^{-2}$ s$^{-1}$ on May 24 and August 7,
respectively. 
Three outbursts were detected at 15 GHz by the Owens Valley Radio Observatory 40-m telescope in 2012 May, 2012
October, and 2013 January, suggesting a complex connection with the
$\gamma$-ray activity. The most likely scenario suggests that the 2012 May
  $\gamma$-ray flare may not be directly related to the radio activity
  observed over the same period, while the two $\gamma$-ray flaring episodes
  may be related to the radio activity observed at 15 GHz in 2012 October and
  2013 January. 
The $\gamma$-ray flare in 2012 May triggered {\em Swift}
observations that confirmed that SBS\,0846$+$513 was also exhibiting high
  activity in the optical, UV and X-ray bands, thus providing a firm identification between the
$\gamma$-ray source and the lower-energy counterpart. We compared the spectral
  energy distribution (SED) of
the flaring state in 2012 May with that of a quiescent state. The two SEDs,
modelled as an external Compton component of seed photons from a dust torus, could be fitted by changing the electron
distribution parameters as well as the magnetic field. No significant evidence
of thermal emission from the accretion disc has been observed. 
Interestingly, in the 5 GHz radio luminosity vs. synchrotron peak frequency plot SBS\,0846$+$513 seems to lie in the flat spectrum radio quasar part of the so-called `blazar sequence'.

\end{abstract}


\section{Introduction}

Narrow-line Seyfert 1 (NLSy1) galaxies are an extreme class of active galactic nuclei
(AGNs) with lower black hole (BH) masses (10$^{6}$--10$^{8}$ M$_{\odot}$) and higher
accretion rates (close to or above the Eddington limit) than those observed in quasars. The strong and
variable radio emission, and the flat radio spectrum suggest the presence of
a relativistic jet in some of them \citep[e.g.][]{yuan08}.  
The detection by the Large Area Telescope (LAT) on-board the {\em Fermi}
satellite of variable $\gamma$-ray emission from 5 radio-loud NLSy1 galaxies brings to three the classes of AGN detected at $\gamma$-ray energies \citep{abdo09,dammando12}, the other two being blazars and radio galaxies. The discovery that NLSy1 host relativistic jets poses intriguing questions about the
nature of these objects, the onset of relativistic jets, the mechanisms of high-energy
emission, and the evolution of radio-loud AGNs. 

\noindent Considering the fact that radio-loud NLSy1s are usually hosted in spiral galaxies \citep[e.g.][]{deo06,zhou06} the presence of a
relativistic jet seems to challenge the firm belief that the formation of
relativistic jets can happen only in elliptical galaxies
\citep{boett02,marscher10}. However, it is worth noting that the NLSy1
Mrk 1239 may be hosted in an early type elliptical/S0 galaxy \citep{mrk89} and
a possible residual of a galaxy merger was observed in 1H 0323$+$342
\citep{anton08}, one of the five NLSy1s detected by {\em Fermi}-LAT.

\noindent One of the key questions is the maximum power released by the jets of radio-loud NLSy1, and for this reason
$\gamma$-ray flaring episodes from these sources have catalyzed a growing
interest in the astrophysical community. The first clues came in 2010 July
and 2011 June when PMN\,J0948$+$0022 underwent a high $\gamma$-ray flaring activity
with daily peak flux integrated above 100 MeV of $\sim$1$\times$10$^{-6}$ ph cm$^{-2}$ s$^{-1}$
\citep{foschini11, dammando11}. Recently, a new $\gamma$-ray flare from this 
source was detected by {\em Fermi}-LAT \citep{dammando13d}. A strong $\gamma$-ray flare was observed also from
SBS\,0846$+$513 in 2011 June--July, reaching an apparent isotropic $\gamma$-ray luminosity of $\sim$10$^{48}$ erg s$
^{-1}$, comparable to that of the bright flat spectrum radio quasars (FSRQs) \citep{dammando12}. This
could be an indication that a few radio-loud NLSy1s are able to host relativistic jets as powerful as those in blazars, despite the lower BH
 masses \citep[e.g.][]{yuan08}.  

The $\gamma$-ray all-sky monitoring by {\em Fermi}-LAT provides us with the
opportunity to follow the daily behaviour of these $\gamma$-ray NLSy1s thus
enabling us to identify flaring episodes. A new period of high $\gamma$-ray activity
from SBS\,0846$+$513 was detected by {\em
  Fermi}-LAT beginning at the end of 2012 April with two peaks in May and
August. 

\noindent The aims of this paper are to discuss the connection between the
  radio and $\gamma$-ray activity of SBS\,0846$+$513, to study the spectral
  energy distribution (SED) of the source in the quiescent and flaring state,
  and to compare the characteristics of this NLSy1 with those of the $\gamma$-ray blazars.
The paper is organized as follows. In Section 2, we
report the LAT data analysis and results. The results of the {\em Swift}
observations are presented in Section 3. Radio data collected by the Owens
  Valley Radio Observatory (OVRO) 40-m, Medicina, and Effelsberg 100-m telescopes
are summarised in Section 4. In Section 5, we present the SED modelling
of the flaring activity in 2012 May and the quiescent state in 2011. Finally, we discuss the radio
variability and proper motion of SBS\,0846$+$513, the connection with the
$\gamma$-ray emission, and draw our conclusions about the comparison with the $\gamma$-ray blazars in Section 6. 

Throughout the paper, the photon index, ${\Gamma}$, is defined as $dN/dE \propto E^{-\Gamma}$, and a $\Lambda$ cold dark matter ($\Lambda$CDM) cosmology
with $H_0$ = 71 km s$^{-1}$ Mpc$^{-1}$, $\Omega_{\Lambda} = 0.73$, and
$\Omega_{\rm m} =
0.27$ is adopted. The corresponding luminosity distance at the source redshift
$z = 0.5835$ \citep{abazajian04} is d$_L = 3.4$\ Gpc, and 1 milliarcsecond corresponds to a
projected linear size of 6.6 pc.

\section{{\em Fermi}-LAT Data: Selection and Analysis}
\label{FermiData}

The {\em Fermi}-LAT  is a pair-conversion telescope operating from 20 MeV to
$>$ 300 GeV. It has a large peak effective area ($\sim$ 8000 cm$^{2}$ for 1
GeV photons), an energy resolution of typically $\sim$10\%, and a field of
view of about 2.4 sr with single-photon angular resolution (68\% containment
radius) of 0.6$^{\circ}$ at {\it E} = 1 GeV on-axis. Further
details about the {\em Fermi}-LAT are given in \citet{atwood09}. 

\begin{figure}
\centering
\includegraphics[width=7.5cm]{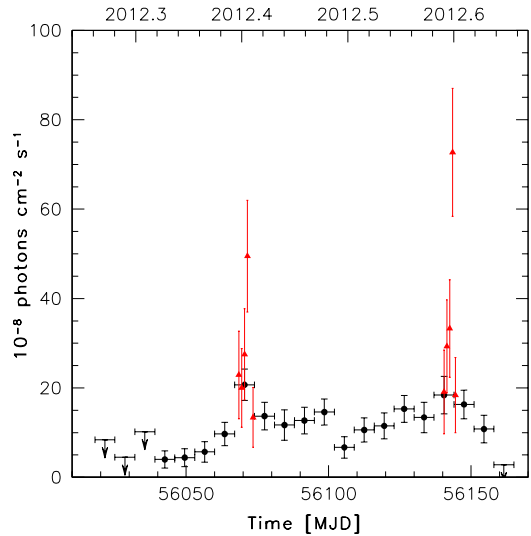}
\caption{Integrated flux light curve of SBS\,0846$+$513 in the 0.1--100 GeV energy range obtained by {\em Fermi}-LAT during 2012 April 1 -- August 28 (MJD 56018--56167) with 7-day or 1-day (shown as triangles) time bins. Arrows refer to 2$\sigma$
  upper limits on the source flux. Upper limits are computed when TS $<$ 10.}
\label{Fig1}
\end{figure}

The LAT data reported in this paper were collected from 2011 December 1 (MJD
55896) to 2013 January 31 (MJD 56323). During this time, the {\em Fermi} observatory operated almost entirely in survey mode. 
The analysis was performed
with the \texttt{ScienceTools} software package version v9r27p1. The LAT data
were extracted within a $10^{\circ}$ region of interest centred at the
radio location of SBS\,0846$+$513. Only events belonging to the `Source' class
were used. The time intervals when the rocking angle of the LAT was greater
than 52$^{\circ}$ were rejected. In addition, a cut on the
zenith angle ($< 100^{\circ}$) was applied to reduce contamination from
the Earth limb $\gamma$ rays, which are produced by cosmic rays interacting with the upper atmosphere. 
The spectral analysis was performed with the instrument response functions
\texttt{P7SOURCE\_V6} using an unbinned maximum-likelihood method implemented 
in the Science tool \texttt{gtlike}. A Galactic diffuse emission model and isotropic component, which is the sum of
an extragalactic and residual cosmic ray background were used to model the background\footnote{http://fermi.gsfc.nasa.gov/ssc/data/access/lat/Background\\Models.html}. The normalizations of both components in the background model were allowed to vary freely during the spectral fitting. 

We evaluated the significance of the $\gamma$-ray signal from the sources by
means of the maximum-likelihood test statistic TS = 2$\Delta$log(likelihood) between models with
and without a point source at the position of SBS\,0846$+$513
\citep{mattox96}. The source model used in \texttt{gtlike} includes all of the point sources from the second {\em Fermi}-LAT catalogue
\citep[2FGL;][]{nolan12} that fall within $20^{\circ}$ of the source. The spectra of these sources were parametrized by power-law functions,
except for 2FGL\,J0920.9$+$4441 for which we used a log-parabola as in the 2FGL
catalogue. A first maximum-likelihood analysis was performed to remove from the model
the sources having TS $<$ 10 and/or the predicted number of counts based on
the fitted model $N_{pred} < 3 $. A second maximum-likelihood analysis
was performed on the updated source model. In the fitting procedure, the normalization factors and the photon indices of
the sources lying within 10$^{\circ}$ of SBS\,0846$+$513 were left as free
parameters. For the sources located between 10$^{\circ}$ and 20$^{\circ}$, we
kept the normalization and the photon index fixed to the values from the 2FGL catalogue.

As already shown in \citet{dammando12}, SBS\,0846$+$513 alternates between periods of low and high $\gamma$-ray activity and therefore was not
  consistently detected in $\gamma$ rays during the {\em Fermi} mission to
  date. In particular, significant $\gamma$-ray activity was observed in
    2011 June--July after a quiescent state extending back at least to the
    start of the {\em Fermi} mission. Integrating over the period 2011 December 1--2012 March 31 (MJD 55896--56017) the fit yielded a TS = 4. The 2$\sigma$ upper
limit is 1.2$\times$10$^{-8}$ ph cm$^{-2}$ s$^{-1}$ in the 0.1--100 GeV
energy range, assuming a photon index of $\Gamma=2.3$. On the contrary, the fit with a power-law model
to the data integrated over the period 2012 April 1--August 31 (MJD 56018--56170) in the 0.1--100 GeV
energy range results in a TS = 851, with an integrated average flux of (12.0 $\pm$ 0.9) $\times$10$^{-8}$ ph cm$^{-2}$ s$^{-1}$ and a
photon index of $\Gamma$ = 2.13 $\pm$ 0.05. Finally, over the period 2012
September 1--2013 January 31 (MJD 56171--56323) a TS of only 10 was obtained, with a
2$\sigma$ upper limit of 2.0$\times$10$^{-8}$ ph cm$^{-2}$ s$^{-1}$ in the 0.1--100 GeV energy range.

In order to test for curvature in the $\gamma$-ray spectrum of SBS\,0846$+$513
during 2012 April--August time frame we used a log-parabola (LP), $dN/dE
\propto$ $E/E_{0}^{-\alpha-\beta \, \log(E/E_0)}$ \citep[]{landau86,
  massaro04}, as an alternative spectral model to the power law (PL). We obtained a spectral slope
$\alpha$ = 1.92 $\pm$ 0.11 at the reference energy $E_0$ = 300 MeV, a curvature parameter around the peak $\beta$ = 0.09 $\pm$
0.04, with a 0.1--100 GeV flux of (11.0 $\pm$ 0.8) $\times$10$^{-8}$ ph cm$^{-2}$
s$^{-1}$ and a TS = 888. We used a likelihood ratio test to check the PL model (null hypothesis) against
the LP model (alternative hypothesis). These values may be compared, following \citet{nolan12}, by defining the curvature test statistic
TS$_{\rm curve}$=(TS$_{\rm LP}$ - TS$_{\rm PL}$)= 37 corresponding to a $\sim$6$\sigma$ difference, indicating a significant curvature in the $\gamma$-ray spectrum of SBS\,0846$+$513 during 2012 April--August. A spectral curvature was also observed during the high activity state in 2011 June \citep{dammando12}. 

Fig.~\ref{Fig1} shows the $\gamma$-ray light curve for the period 2012 April
1--August 28 using a log-parabola model and 1-week time bins. For the highest
significance periods we also reported fluxes in 1-day time intervals. For each time bin, the spectral parameters for
SBS\,0846$+$513 and for all the sources within 10$^{\circ}$ from it were frozen to
the value resulting from the likelihood analysis over the entire period. If TS
$<$ 10, 2$\sigma$ upper limits were evaluated. The systematic
uncertainty in the flux measurement is energy dependent: it amounts to $10\%$ at 100 MeV, decreasing to
$5\%$ at 560 MeV, and increasing to $10\%$ above 10 GeV \citep{ackermann12}.

The first emission peak was observed on 2012 May 24 (MJD 56071), with an
average flux for that day of (50 $\pm$ 12)$\times$10$^{-8}$ ph cm$^{-2}$
s$^{-1}$ in the 0.1--100 GeV energy range, corresponding to an
apparent isotropic $\gamma$-ray luminosity of $\sim$5 $\times$10$^{47}$ erg
s$^{-1}$, a factor of 2 lower than the peak value observed in 2011
June \citep{dammando12}. A second peak at higher flux with respect to the
first one, (73 $\pm$ 14)$\times$10$^{-8}$ ph cm$^{-2}$ s$^{-1}$
(corresponding to an apparent isotropic $\gamma$-ray luminosity of $\sim$8
$\times$10$^{47}$ erg s$^{-1}$), was detected on 2012 August 7 (MJD
56146). In both of the flares, a doubling time scale of 1-2 days was observed.  By means of the \texttt{gtsrcprob} tool we estimated that
the highest energy photon detected from SBS\,0846$+$513 was observed on 2012
August 8 at a distance of 0.08$^{\circ}$ from the source with an energy of 16.1
GeV. Interestingly, this highest energy photon was detected at the
peak of the 2012 flaring activity.
 
\begin{table*}
\caption{Log and fitting results of {\em Swift}/XRT observations of
  SBS\,0846$+$513 using a power-law model with a HI column density fixed to the
  Galactic value in the direction of the source. $^{a}$Observed flux.}
\begin{center}
\begin{tabular}{cccccc}
\hline \hline
\multicolumn{1}{c}{Date} &
\multicolumn{1}{c}{Date} &
\multicolumn{1}{c}{Net exposure time} &
\multicolumn{1}{c}{Net count rate} &
\multicolumn{1}{c}{Photon index} &
\multicolumn{1}{c}{Flux 0.3--10 keV$^{a}$} \\
\multicolumn{1}{c}{(MJD)} &
\multicolumn{1}{c}{(UT)} &
\multicolumn{1}{c}{(sec)} &
\multicolumn{1}{c}{($\times$10$^{-2}$ cps)} &
\multicolumn{1}{c}{($\Gamma$)} &
\multicolumn{1}{c}{($\times$10$^{-13}$ erg cm$^{-2}$ s$^{-1}$)} \\
\hline
55915-55929 & 2011-12-20/2012-01-03 & 6291 & 1.7 $\pm$ 0.2 & $1.6 \pm 0.2$ & $8.8 \pm 1.3$ \\
56075 & 2012-05-27 & 1968 &  2.4 $\pm$ 0.3 & $1.6 \pm 0.3$ & $12.2 \pm 2.7$ \\
56085 & 2012-06-07 & 2782 & 1.5 $\pm$ 0.2 & $1.5 \pm 0.3$ & $9.0 \pm 2.1$ \\
56230 & 2012-10-30 & 4815 & 0.7 $\pm$ 0.1 & $1.6 \pm 0.4$ & $4.0 \pm 0.8$ \\
56261 & 2012-11-30 & 4815 & 0.7 $\pm$ 0.1 & $1.7 \pm 0.4$ & $3.3 \pm 0.6$ \\
56291 & 2012-12-30 & 4810 & 0.7 $\pm$ 0.1 & $1.6 \pm 0.4$ & $3.7 \pm 0.9$ \\
56322 & 2013-01-30 & 4700 & 1.2 $\pm$ 0.2 & $1.8 \pm 0.3$ & $5.8 \pm 0.7$ \\
\hline
\hline
\end{tabular}
\end{center}
\label{XRT}
\end{table*}

\section{{\em Swift} Data: Analysis and Results}
\label{SwiftData}

The {\em Swift} satellite \citep{gehrels04} performed ten observations
of SBS\,0846$+$513 between 2011 December and 2013 January. The observations were
performed with all three on-board instruments: the X-ray Telescope \citep[XRT;][0.2--10.0 keV]{burrows05}, the Ultraviolet/Optical Telescope \citep[UVOT;][170--600 nm]{roming05} and the Burst Alert Telescope \citep[BAT;][15--150 keV]{barthelmy05}.

The hard X-ray flux of this source is below the sensitivity of the BAT
instrument for the short exposure of these observations and therefore the data from this instrument are not used.
Moreover, the source was not present in the {\em Swift} BAT 70-month hard X-ray catalogue \citep{baumgartner13}.

The XRT data were processed with standard procedures (\texttt{xrtpipeline v0.12.6}), filtering, and screening criteria using the \texttt{HEAsoft} package
(v6.12). The data were collected in photon counting mode for all of the observations. The source count rate
was low ($<$ 0.5 counts s$^{-1}$); thus pile-up correction was not
required. The data collected during the four observations performed between 2011 December 20 and 2012 January 3 were
summed in order to have enough statistics to obtain a
good spectral fit. Source events were extracted from a circular region with a radius of
20 pixels (1 pixel $\sim$ 2.36$"$), while background events were extracted from a circular region with radius of 50 pixels away from the source
region. Ancillary response files were generated with \texttt{xrtmkarf}, and
account for different extraction regions, vignetting and point-spread function
corrections. We used the spectral redistribution matrices v013 in the
Calibration data base maintained by HEASARC\footnote{http://heasarc.nasa.gov/}. Considering the low number of photons collected ($<$ 200 counts) the spectra
were rebinned with a minimum of 1 count per bin and we used the Cash statistic
\citep{cash79}. We fitted the spectrum with an absorbed power-law using the photoelectric absorption model
\texttt{tbabs} \citep{wilms00}, with a neutral hydrogen column density fixed
to its Galactic value \citep[2.9$\times$10$^{20}$cm$^{-2}$;][]{kalberla05}.
The fit results are reported in 
Table~\ref{XRT}. As was seen in previous XRT observations
\citep[see][]{dammando12} the X-ray  
spectrum of SBS\,0846$+$513 is harder than the other NLSy1s \citep[$\Gamma_{\rm X} \, > 2$, e.g.][]{grupe10},
suggesting a significant contribution of inverse Compton radiation from a
relativistic jet, similar to for FSRQs. We noted an increase of $\sim40\%$
in the net count rate (and flux) observed on 2012 May 27 with respect to 2011 December--2012
January and up to a factor of $\sim$3 with respect to 2012 October--December, in
agreement with the increase of the $\gamma$-ray activity observed by the {\em
  Fermi}-LAT on the same days. Unfortunately no simultaneous {\em Swift}
observations are available during the 2012 August $\gamma$-ray flare. No
significant change of the photon index was observed during the entire year 2012.

UVOT data in the $v$, $b$, $u$, $w1$, $m2$, and $w2$ filters were reduced with the \texttt{HEAsoft} package v6.12 and the 20120606 CALDB-UVOTA
release. We extracted the source counts from a circle with 5 arcsec radius centred
on the source and the background counts from a circle with 10 arcsec radius in a
near, source-free region. 
As in \citet{dammando12}, we calculated the effective wavelengths, count-to-flux conversion factors, and amount of Galactic extinction in the UVOT
bands by convolving the physical quantities with a power-law fit to the source flux
and with the filter effective areas. The results are shown in Table~\ref{caluvot}. The differences with respect to  \citet{dammando12} are due to the new versions
of both software and calibration files, as well as to the re-calibration of dust
reddening by \citet{schlafly11}. For this reason we also re-analyse the observation performed on 2011 September 15, already presented in \citet{dammando12}. 

\begin{table}
\caption{Results of the UVOT calibration procedure: effective wavelengths
  $\lambda_{\rm eff}$, count rate to flux conversion factors $\rm CF_\Lambda$,
  and Galactic extinction calculated from the \citet{cardelli89} laws.}     
\centering                          
\begin{tabular}{l c c c}       
\hline\hline                 
Filter & $\lambda_{\rm eff}$ & $\rm CF_\Lambda$ & $A_\Lambda$ \\    
       & (\AA) & ($10^{-16} \rm erg \, cm^{-2} s^{-1}$ \AA$^{-1}$) & (mag) \\
\hline                        
   $v$    & 5437 & 2.60 & 0.072 \\      
   $b$    & 4374 & 1.47 & 0.095 \\
   $u$    & 3489 & 1.65 & 0.110 \\
   $uvw1$ & 2676 & 4.40 & 0.160 \\
   $uvm2$ & 2270 & 8.35 & 0.200 \\ 
   $uvw2$ & 2136 & 5.99 & 0.190 \\ 
\hline  
\hline                                
\end{tabular}
\label{caluvot} 
\end{table} 

\begin{figure}
\centering
\includegraphics[width=7.5cm]{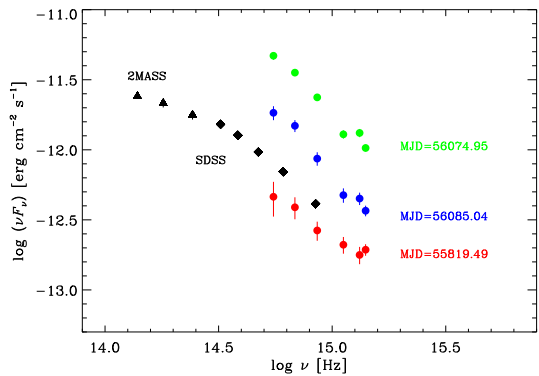}
\caption{Spectral energy distributions of SBS\,0846$+$513 collected in
  optical/UV by {\em Swift}/UVOT on 2011 September 15 (MJD 55819), 2012 May 27
  (MJD 56074), and June 7 (MJD 56085). In addition archival infrared 2MASS and
  optical SDSS data, collected on 1999 December 23 and 2000 November 23
  respectively, are shown.}
\label{UVOT_SED}
\end{figure}

The UVOT photometry is reported in Table~\ref{UVOT}. We note a significant
increase of more than 2 mag (a factor of $\sim$10 in flux density) in
all of the UVOT filters on 2012 May 27 with respect to the observations performed on 2012 January 3. This
high optical/UV activity is followed by a decrease of $\sim$1 mag (a factor of $\sim$2.5 in flux density) in 10 days. On October 30 the source returned to a low state, comparable to the
level observed in 2012 January. These large variations in optical/UV are most
likely due to a variation in the jet emission.  
In Fig.~\ref{UVOT_SED} we display the SEDs from 2012 May
and June as well as the one from 2011 September, which was already presented in
\citet{dammando12}. Archival Two Micron All Sky Survey (2MASS) and Sloan
Digital Sky Survey (SDSS) data, collected on 1999 December 23 and 2000
November 23 respectively, are also shown. The comparison between the
SEDs collected during different activity states confirms a significant
increase in the synchrotron emission on 2012 May 27, soon after the first peak of the
$\gamma$-ray activity, together with a small UV bump. We investigate the
  possibility that this is a hint of the contribution of the accretion disc emission.
The signature of the accretion disc has been observed in FSRQs during both low
activity states \citep[e.g.~3C 454.3;][]{raiteri11} and high activity states
\citep[e.g.~PKS\,1510$-$089;][]{dammando11b}, but in the second case that signature
was also well detected during the low state. Therefore, the lack of this feature
during the low state of SBS\,0846$+$513 in 2011 makes it unlikely that we are
detecting accretion disc emission.

\begin{table*}
\caption{Results of the {\em Swift}/UVOT observations of SBS\,0846$+$513 in magnitudes. Upper limits
are calculated when the analysis provided a significance of detection $<$3$\sigma$.}
\begin{center}
\begin{tabular}{cccccccc}
\hline \hline
\multicolumn{1}{c}{Date (MJD)} &
\multicolumn{1}{c}{Date (UT)} &
\multicolumn{1}{c}{$v$} &
\multicolumn{1}{c}{$b$} &
\multicolumn{1}{c}{$u$} &
\multicolumn{1}{c}{$w1$} &
\multicolumn{1}{c}{$m2$} &
\multicolumn{1}{c}{$w2$} \\
\hline
55819 & 2011-09-15 & 19.18$\pm$0.30 & 19.75$\pm$0.19 & 19.29$\pm$0.17 & 19.47$\pm$0.15  & 19.62$\pm$0.15 & 19.62$\pm$0.10\\
55915 & 2011-12-20 & $>$ 18.81 &  $>$ 19.89 & $>$ 19.57 & 20.43$\pm$0.27 & $>$ 19.69 & 20.14$\pm$0.31 \\
55922 & 2011-12-27 & -- & $>$ 20.11 & 19.65$\pm$0.33 & $>$20.11 & 19.78$\pm$0.21 & 20.48$\pm$0.38 \\
55923 & 2011-12-28 & $>$ 18.88 &  $>$ 19.84 & 19.42$\pm$0.36 & 20.14$\pm$0.16 & $>$ 19.83 & 20.13$\pm$0.29 \\
55929 & 2012--01-03 & -- & -- & -- & -- & -- & $>$20.42 \\
56074 & 2012-05-27 & 16.66$\pm$0.07 &  17.34$\pm$0.06 & 16.92$\pm$0.06 & 17.50$\pm$0.07 & 17.44$\pm$0.03 & 17.80$\pm$0.06 \\
56085 & 2012-06-07 & 17.68$\pm$0.12 &  18.30$\pm$0.11 & 18.01$\pm$0.12 & 18.59$\pm$0.13 & 18.62$\pm$0.11 & 18.92$\pm$0.09 \\
56230 & 2012-10-30 & $>$ 19.18 &  $>$ 20.21 & 19.85$\pm$0.38 & 19.69$\pm$0.26 & 19.98$\pm$0.11 & 20.36$\pm$0.29 \\
56261 & 2012-11-30 & $>$ 18.86 &  19.78$\pm$0.28 & $>$19.81 & 19.88$\pm$0.32 & 19.56$\pm$0.26 & 20.22$\pm$0.16 \\
56291 & 2012-12-30 & 19.01$\pm$0.27 & 20.27$\pm$0.34 & $>$20.03 & 20.06$\pm$0.13 & 20.19$\pm$0.31 & 20.45$\pm$0.25 \\
56322 & 2013-01-30 & 19.30$\pm$0.35 & $>$20.32 & $>$19.97 & $>$20.31 & 20.11$\pm$0.29 & 20.55$\pm$0.28 \\ 
\hline
\hline
\end{tabular}
\end{center}
\label{UVOT}
\end{table*}

\section{Radio Data: Analysis and Results}\label{RadioData}

\subsection{OVRO}

\begin{figure}
\centering
\includegraphics[width=7.5cm]{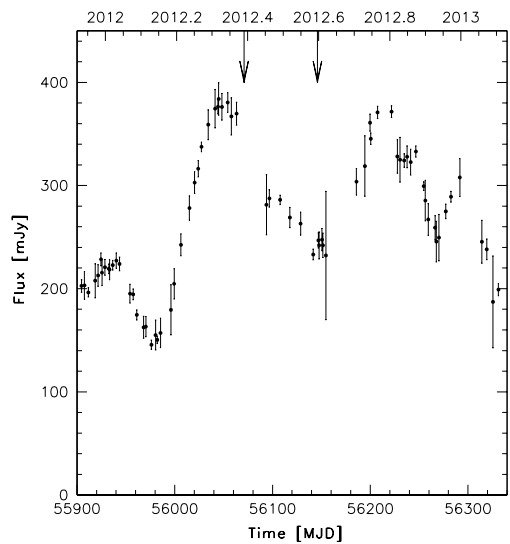}
\caption{15 GHz radio light curve of SBS\,0846$+$513 for the period 2011 December--2013 January from the OVRO 40-m telescope. The two downward
arrows indicate the times of the peaks of the $\gamma$-ray flaring episodes.}
\label{Fig3}
\end{figure}

As part of an ongoing blazar monitoring programme, the OVRO 40-m radio telescope has observed SBS\,0846$+$513
at 15~GHz regularly since the end of 2007 \citep{richards11}. This
monitoring programme includes about 1700 known and likely $\gamma$-ray loud
blazars above declination $-20^{\circ}$. The sources in this
program are observed in total intensity twice a week with a 4~mJy
(minimum) and 3\% (typical) uncertainty on flux densities. Observations are performed
with a dual-beam (each 2.5~arcmin FWHM) Dicke-switched system using
cold sky in the off-source beam as the reference. Additionally, the
source is switched between beams to reduce atmospheric variations. The
absolute flux density scale is calibrated using observations of
3C~286, adopting the flux density (3.44~Jy) from \citet{baars77}. 
This results in about a 5\% absolute scale uncertainty, which is not reflected
in the plotted errors. SBS\,0846$+$513 was highly variable at 15 GHz during the
OVRO 40-m telescope monitoring (Fig.~\ref{Fig3}), with a flux density rising
from 145 mJy (at MJD 55976) to 384 mJy (at MJD 56044).

\subsection{Medicina and Effelsberg 100-m}

SBS\,0846$+$513 was observed at 8.4 GHz with the Medicina radio telescope on 2012 June 18 (MJD 56096), August 2 (MJD 56141), and October 4 (MJD 56204). The new Enhanced Single-dish Control
System (ESCS) acquisition system, which provides increased sensitivity and supports observations with the cross scan technique was used.
The typical on-source time was 1.5 minutes and the flux density was
calibrated with respect to 3C\,286, 3C\,48, and NGC\,7027. Since the
signal-to-noise ratio in each scan across the source was low (typically
$\sim3$), a stacking analysis of the scans was performed. A slight
  increase of the 8.4 GHz flux density from 0.20$\pm$0.02 Jy to 0.26$\pm$0.02 Jy was
  observed between 2012 June and October. The observation on October 4
  was carried out also at 5 GHz and, together with the OVRO data collected at
  15 GHz on October 7, a spectral index of $-$0.6$\pm$0.2 was estimated between 5 and 15 GHz.

\begin{figure}
\centering
\includegraphics[width=7.5cm]{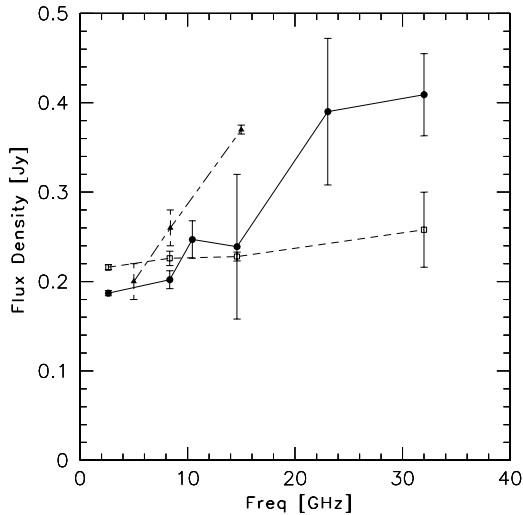}
\caption{Radio spectra of SBS\,0846$+$513 obtained by Effelsberg on 2011
    April 30 (MJD 55681; open squares), 2012 August 20 (MJD 56159; filled
    circles), and Medicina and OVRO on 2012 October 4--7 (MJD 56204--56207;
    filled triangles).}
\label{effelsberg}
\end{figure}

The radio spectra of SBS\,0846$+$513 were also observed with the Effelsberg 100-m
telescope from 2.64 GHz to 32 GHz on 2012 July 1 (MJD 56109), August 5 (MJD 56144), and August 20 (MJD
56159) within the framework of a {\em Fermi}-related monitoring programme of
$\gamma$-ray blazars \citep[F-GAMMA programme;][]{fuhrmann07,angelakis10}. Further
  details about observation mode and data reduction  are reported in
  \citet{fuhrmann08} and \citet{angelakis08}. The flux density at 32
  GHz increased from 0.32$\pm$0.03 Jy on 2012 July 1 to 0.41$\pm$0.05 Jy on August 20, consistent with the behaviour observed by OVRO at 15 GHz.
In Fig.~\ref{effelsberg} we compare the radio spectrum collected by Effelsberg
on 2011 April 30 \citep[MJD 55681; already presented in][]{dammando12} to the
spectra collected by Effelsberg on 2012 August 20 (MJD 56159), and OVRO and
Medicina on 2012 October 4--7 (MJD 56204--56207). See the discussion in
Section \ref{comparison}.

\section{SED modeling}
\label{SED_modeling}

\begin{figure} 
\centering
\vspace{2.0mm}
\includegraphics[width=7.5cm]{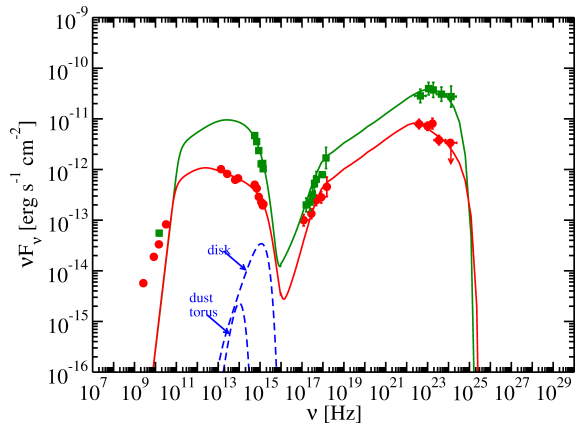}
\caption{Spectral energy distribution data (squares) and model fit
  (solid curve) of SBS\,0846$+$513 in flaring activity with the model components shown as dashed
  curves. The data points were collected by OVRO 40-m (2012 May 17), {\em Swift}
  (UVOT and XRT; 2012 May 27), and {\em Fermi}-LAT (2012 May 20--29). The SED
  in the quiescent state reported in \citet{dammando12} is shown as circles.}
\label{0846sed}
\end{figure}

The lack of multi-wavelength data simultaneous with the 2011 June--July $\gamma$-ray
flare observed by LAT did not allow us to investigate the
SED of this source during a high state in \citet{dammando12}. Thanks to the {\em Swift} observations
performed in 2012 May, soon after the new $\gamma$-ray flare, we are now able to characterize the
flaring state of the source and compare it with a quiescent state.
In Fig.~\ref{0846sed}, we plot the SED during a quiescent state in 2011 from~\citet{dammando12}, and a simultaneous SED based on the
flare in 2012 May. The flaring state includes the LAT spectrum built with data
centred on 2012 May 20 to 29 (MJD 56067--56076), the {\em Swift} (UVOT
and XRT) data collected on 2012 May 27 (MJD 56074) and the OVRO 40-m data closest to the $\gamma$-ray peak, collected on 2012 May 17 (MJD 56064). In addition to the quiescent state presented in \citet{dammando12}, we
modelled the 2012 May flaring state with a combination of synchrotron,
synchrotron self-Compton (SSC), and external Compton from dust torus
emission. The description of the model can be found in \citet{finke08_ssc} and \citet{dermer09} and jet powers were
calculated assuming a two-sided jet.
The modelling results are presented in Fig.~\ref{0846sed}
and Table~\ref{table_fit}. For some FSRQs, it has been found that the
fits to the flaring and quiescent states can be made by varying just the
electron distribution, while keeping the other parameters fixed
\citep[e.g.~PKS\,0537$-$441,][]{dammando13}. Our first attempts to fit the 2012 flaring state by varying only the electron distribution parameters
failed. We subsequently adjusted the dust torus parameters, the Lorentz
factor, and the magnetic field in order to reproduce the flaring state.  We then
attempted to reproduce the 2011 quiescent state by varying only the
electron distribution parameters from the flaring state fit. These
attempts also failed. Finally, we found that by varying the magnetic
field as well as the electron distribution parameters we could
reasonably model the quiescent state. It has been suggested that some
flares from FSRQs require a change in the magnetic field strength
\citep[e.g.~PKS\,0208$-$512,][]{chatterjee13}, and this result is consistent with that
suggestion.

The dust component used in this paper is significantly fainter than that
  used for the SED modeling of the quiescent state in \citet{dammando12}, but
  it is compensated for by the higher Lorentz factor.  Both the quiescent and
flaring state models shown in Table~\ref{table_fit} are near
equipartition between the electron and the magnetic field energy
densities. We noted a significant shift of the synchrotron peak ($\nu_{pk}^{sy}$) to higher frequencies from $\nu_{pk}^{sy}$
  $\sim$2$\times$10$^{12}$ Hz during the quiescent state to $\sim$ 2$\times$10$^{13}$ Hz during the flaring state, similar to FSRQs
\citep[e.g.~PKS\,1510$-$089,][]{dammando11b}. The flaring state has higher jet powers in terms of both electrons
and magnetic field.  Also, for both the flaring and quiescent states,
a super-exponential cutoff at high energies in the electron distribution ($N_e \propto \gamma^{\prime - p_2} \exp[-(\gamma/\gamma_2)^4]$) was
required so that the observed SEDs could be reasonably reproduced. If the jet is conical and the emitting region takes up the entire cross section of
the jet, the jet half-opening angle will be $\alpha \approx
R_b^\prime/r \approx 2.7^\circ$, consistent with the opening angles
measured from VLBI for many blazars \citep[e.g.][]{jorstad05}. In Fig.~\ref{0846sed}, the
blue bump from the accretion disc is also plotted, assuming a conservative
value for the BH mass of $10^8\ M_\odot$ \citep[see the discussion on the BH mass of this source in][]{dammando12} and a luminosity about 10 times that of the dust
torus. The disc luminosity is radiating at
$L_{disk}/L_{Edd} \approx 3\times10^{-3}$ of the Eddington luminosity, quite a low value.  This
value is constrained by the lack of a blue bump observed from this
source in the quiescent state, and could not, therefore, be significantly higher. 

\noindent The considered synchrotron component is self-absorbed below
$\sim$10$^{11}$ Hz. A larger radius of the blob is necessary to also fit
  the radio data, but this is not compatible with the rapid variability observed in $\gamma$ rays during the flaring
state. An alternative solution is that a small compact region responsible
for the $\gamma$-ray flare is nearly co-spatial with the
region producing the radio outburst, as proposed e.g. by \citet{marscher10c}.

\begin{table*}
\footnotesize
\begin{center}
\caption{Model parameters for the SED shown in Fig.~\ref{0846sed}.
\label{table_fit}}
\begin{tabular}{lccc}
\hline \hline
Parameter & Symbol & 2011 Quiescent State & 2012 Flaring State \\
\hline
Redshift & 	$z$	& 0.5835	& 0.5835	  \\
Bulk Lorentz Factor & $\Gamma$	& 40	& 40	  \\
Magnetic Field [G]\tablenotemark{**} & $B$         & 0.20   & 0.25    \\
Variability Time-Scale [s]& $t_v$       & 1$\times$$10^5$ & 1$\times$$10^5$  \\
Comoving radius of blob [cm]& R$^{\prime}_b$ & 7.6$\times$10$^{16}$ & 7.6$\times$10$^{16}$ \\
Jet Height [cm]& $r$ & $1.6\times10^{18}$ & $1.6\times10^{18}$ \\
Low-Energy Electron Spectral Index & $p_1$       & 2.4 & 2.4     \\
High-Energy Electron Spectral Index\tablenotemark{**}  & $p_2$       & 3.3 & 3.0	 \\
Minimum Electron Lorentz Factor & $\gamma^{\prime}_{min}$  & $1.3$ & $1.3$ \\
Break Electron Lorentz Factor\tablenotemark{**} & $\gamma^{\prime}_{brk}$ & $3.0\times10^2$ & $6.7\times10^2$ \\
Maximum Electron Lorentz Factor\tablenotemark{**} & $\gamma^{\prime}_{max}$  & $8.0\times10^3$ & $5.0\times10^3$ \\
Dust Torus luminosity [erg s$^{-1}$] & $L_{dust}$ & $4.4\times10^{42}$ & $4.4\times10^{42}$   \\
Dust Torus temperature [K] & $T_{dust}$ & $2.0\times10^3$ & $2.0\times10^3$ \\
Dust Torus radius [cm] & $R_{dust}$ & $4.1\times10^{18}$ & $4.1\times10^{18}$  \\
Jet Power in Magnetic Field [erg s$^{-1}$] & $P_{j,B}$ & $2.7\times10^{45}$ & $4.3\times10^{45}$  \\
Jet Power in Electrons [erg s$^{-1}$] & $P_{j,par}$ & $1.7\times10^{45}$ & $5.5\times10^{45}$ \\
\hline \hline
\end{tabular}
\end{center}
\tablenotetext{**}{Parameters changed between the quiescent and the flaring state.}
\end{table*}

\section{Discussion and conclusions}
\label{discussion}

After the spatial association between the $\gamma$-ray source and
the counterpart at lower energies presented in \citet{dammando12}, the
significant increase of activity detected by {\em Swift} and {\em Fermi} almost simultaneously in the optical, UV, X-ray
and $\gamma$-ray bands in 2012 May (see Fig.~\ref{MWL}) has enabled us to firmly identify the $\gamma$-ray source with the NLSy1 SBS\,0846$+$513. The radio-to-$\gamma$-ray data collected between 2011 December and 2013 January allowed us also to investigate some properties of this source. 

\subsection{Radio variability and proper motion}

\begin{figure} 
\centering
\vspace{2.0mm}
\includegraphics[width=7.0cm]{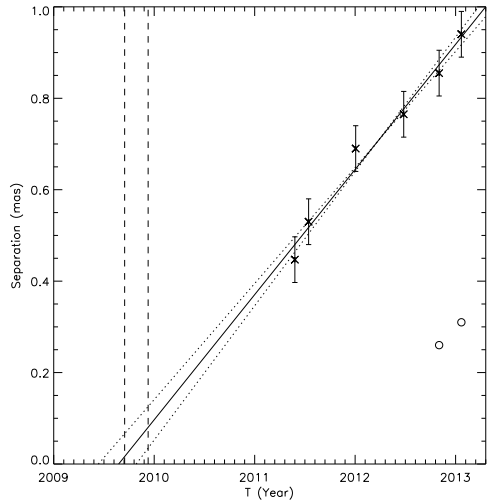}
\caption{The separation between the core component of SBS\,0846$+$513 and the knot ejected in 2009 as a function of time. The solid line represents the regression fit to the 15 GHz VLBA MOJAVE data, while the dotted lines represent the uncertainties from the fit parameters. Two dashed vertical lines indicate the beginning and the peak of the radio flare observed by OVRO 40-m. The first two epochs results for the knot likely ejected in 2011 are also reported as open circles.}
\label{regression}
\end{figure}

Contrary to what was observed in 2011 during the first $\gamma$-ray flaring event of SBS\,0846$+$513 detected by {\em Fermi}-LAT
\citep{dammando12}, three outbursts are clearly seen in the 15 GHz light curve, between 2012 May
and 2013 January (see Figs.~\ref{Fig3} and \ref{MWL}). These data indicate a significant increase of
the flux density at 15 GHz starting close in time to the 2012 May
$\gamma$-ray flare. In particular, an increase by a factor of $\sim$2.5 was observed between 2012 February 19 (MJD 55976) and 2012 May 7 (MJD
56054). The peak flux density of 384 mJy reached
during the outburst is the highest value observed at 15 GHz for this source by the OVRO 40-m
telescope since 2011 January. Previously, at the end of 2009, a strong radio flare was detected at 15 GHz, reaching a
peak flux density of 517 mJy on 2009 December 9
\citep[see][]{dammando12}. Following the procedure described in
\citet{orienti13} the analysis of the 6-epoch data-set collected by the Monitoring Of Jets in Active galactic nuclei with VLBA Experiments (MOJAVE) programme\footnote{The MOJAVE data archive is maintained at http://www.physics.purdue.edu/MOJAVE} during 2011--2013 indicates that a
superluminal jet component is moving away from the core with an apparent
angular velocity of
(0.27$\pm$0.02) mas yr$^{-1}$, which corresponds to (9.3$\pm$0.6)$c$ (Fig.~\ref{regression}). This velocity
estimate is more
accurate than the value derived in \citet{dammando12b} thanks to the larger
number of observing epochs available spanning a longer time interval.
The time of zero separation estimated by means of a linear regression fit is $T_{0}$ = 2009.646 (2009 August 24). This suggests a strict connection
between the 2009 radio flare and the ejection of this component.  
However, the uncertainty in the velocity does not allow us to accurately constrain the precise time of zero separation,
which could be between 2009.445 (i.e.~2009 June 11) and 2009.821 (i.e.~2009
October 27) (Fig.~\ref{regression}). Interestingly, no
significant $\gamma$-ray activity from SBS\,0846$+$513 was detected
simultaneously to the radio outburst and the ejection of this knot \citep[see][]{dammando12}. This is 
different from what is observed in many bright
blazars like PKS\,1510$-$089, where $\gamma$-ray flares occur close in time with the ejection of
superluminal knots and an increase of both the total and polarized radio flux
\citep[see e.g.][]{marscher10b, marscher12, orienti13}. On the contrary, a
tentative detection of a new feature at about 0.25 mas from the core was obtained in the last
two epochs of MOJAVE observations. This new component is separating from the core with an angular velocity of 0.23 mas yr$^{-1}$,
corresponding to an apparent velocity of 7.7$c$. This gives a time of zero-separation of 2011.70
(2011 September 12). The new feature should have been ejected close in time
with the $\gamma$-ray flare in 2011 June--July. The availability of only
  two observing epochs, and the large uncertainties on the blob position do
  not allow us to confirm this result so far. If this new component is
  confirmed by further MOJAVE epochs, it will be an indication of a different relation between the knot ejection and the $\gamma$-ray activity for SBS\,0846$+$513 in 2009 and 2011.

\subsection{Radio and $\gamma$-ray connection in 2012}

\begin{figure*}
\centering
\includegraphics[width=11cm]{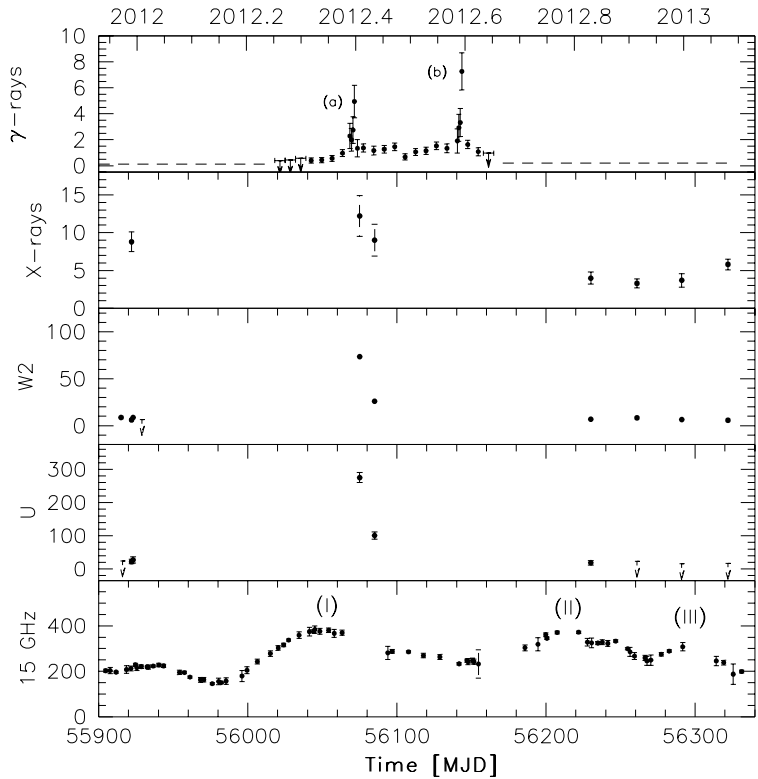}
\caption{Multifrequency light curve for SBS\,0846$+$513. The period covered is 2011 December--2013 January. The data sets were collected (from top to
  bottom) by {\em Fermi}-LAT ($\gamma$ rays, 0.1--100 GeV; in units of 10$^{-7}$ ph
  cm$^{-2}$ s$^{-1}$), {\em Swift}-XRT (0.3--10 keV; in units of 10$^{-13}$ erg cm$^{-2}$ s$^{-1}$), {\em Swift}-UVOT ($w2$ and $u$
bands; in units of $\mu$Jy), and OVRO (15 GHz; in units of mJy). Arrows
refer to 3$\sigma$ upper limits on the source flux densities for the $w2$
  and $u$ bands, and to 2$\sigma$ upper limits on the source fluxes for the
  $\gamma$-ray light curve. During 2011 December--2012 April and 2012 September--2013 January upper
  limits of 1.2$\times$10$^{-8}$ and 2.0$\times$10$^{-8}$ ph cm$^{-2}$
  s$^{-1}$ are shown as dashed lines in the $\gamma$-ray light curve. The $\gamma$-ray
and radio peaks are labelled with (a) and (b), and (I), (II), and (III),
respectively. 
}
\label{MWL}
\end{figure*}

The $\gamma$-ray activity observed in 2011 and 2012 for SBS\,0846$+$513 could be related to different flaring episodes. No significant $\gamma$-ray
  emission was detected during 2011 December--2012 March (see Fig.~\ref{MWL})
  or during 2011 September--November \citep{dammando12}. It is unlikely that
  the 2012 May radio flare is related to the 2011 June $\gamma$-ray flare
  previously observed by {\em Fermi}-LAT. Therefore, a different connection
  between radio and $\gamma$-ray emission is possible for the two high
  activity periods, as has been observed for some blazars \citep[e.g. PKS\,1510$-$089;][]{orienti11,orienti13}. In the case of the high activity observed in 2012
May, we first consider a
common origin for the radio and $\gamma$-ray emission (with the $\gamma$-ray and radio peaks labelled with (a) and (I), respectively, in Fig.~\ref{MWL}). The flux increase in these bands occurs close in time,
suggesting that the radio and $\gamma$-ray emission may originate in the same
part of the jet, likely at large distance from the central engine, where the opacity
effects should be less severe.
No new superluminal component has been detected so far that is likely related to this
flare. Further high spatial resolution observations
of the source are necessary to detect a possible new component ejected at that
period. A second radio peak was observed on 2012 October 7 (MJD 56207; labelled with (II) in Fig.~\ref{MWL}) with a
similar flux density of that observed in May. This peak is delayed by $\sim$2 months with
respect to the $\gamma$-ray flare observed on 2012 August 7 (MJD 56146; labelled with (b) in Fig.~\ref{MWL}). This
could be an indication that the second $\gamma$-ray flaring activity was
produced close to the BH, where the opacity effects should be more severe, causing a time delay at radio wavelengths. 

The presence of a third radio outburst during 2012 December--2013 January peaking on 2012
December 30 (MJD 56291; labelled with (III) in Fig.~\ref{MWL}) suggests an
alternative scenario where the first $\gamma$-ray flare in 2012 May is not
directly connected with the radio activity observed by OVRO 40-m in the same period. In this
case, the 2012 May and August $\gamma$-ray flaring episodes are related to the radio
activity observed at 15 GHz on 2012 October and 2013 January, respectively. In
both cases the peak of the
radio emission is delayed with respect to the $\gamma$-ray one by $\sim$4.5
months. This delay may be due to opacity in the core region of the source at
radio frequencies, suggesting that the $\gamma$-ray emission is produced
relatively close to the BH in both the 2012 May and August $\gamma$-ray
flares. On the basis of the delay between $\gamma$-ray
emission and radio emission at 15 GHz usually observed in blazar-like objects
\citep[see e.g.][]{pushkarev10}, this second
scenario seems to be favoured. Taking into account the lack of
  significant $\gamma$-ray emission from SBS\,0846$+$513 between 2011
  December and 2012 April, it is difficult to associate the radio activity observed in 2012 May 
  with a previous $\gamma$-ray flare. This further suggests a complex relation between the
  radio and $\gamma$-ray emission in this source. 

The 15 GHz light curve is highly variable (see Fig.~\ref{Fig3}). For this
reason we can estimate the variability Doppler factor on the basis of the
radio data collected during an outburst in a similar manner to that
  proposed by \citet{valtoja99}. Given the frequent time sampling during
  the first part of the OVRO light curve we can estimate the rise time
  ($\Delta t$) of the first flaring episode. However, the second and third
  outbursts were not well sampled by the observations, precluding a reliable determination
of its variability time-scale. We consider $\Delta t$ as the time
  interval of the flux density variation between the maximum
  and minimum flux density of a single outburst $\Delta S$.
 This assumption implies that the minimum flux density
  corresponds to a stationary underlying component and that the variation is due to a
  transient component. On the basis of the causality argument we can derive
  the brightness temperature and then the variability Doppler factor.
 Following \citet{dammando13b} we compute the variability brightness temperature by means of 

\begin{equation}
T_{B,\rm var} = \frac{2}{\pi k} \frac{\Delta S d_{L}^{2}}{\Delta t^{2} \nu^{2}
  {\rm (} 1+z {\rm )^{1+ \alpha}}},
\label{tbright}
\end{equation}

\noindent where $k$ is the Boltzmann constant, $\nu$ is the observing
frequency, and $\alpha$ is the spectral index. In the case of the first outburst we have $\Delta t =
62$ d and $\Delta S=234$ mJy. If in equation~\ref{tbright} we consider these
values and we assume $\alpha =0$, we obtain $T_{B,\rm var} =
1.14\times 10^{14}$ K, which is much larger than the value derived for the
Compton catastrophe \citep[see e.g.][]{kellermann69}. Assuming
that such a high value is due to Doppler boosting, we can estimate the
variability Doppler factor $\delta_{\rm var}$, by means of

\begin{equation}
\delta_{var} = \left( \frac{T_{B,\rm var}}{T_{B,\rm int}} \right)^{1/(3+ \alpha)},
\label{dopplervar}
\end{equation}

\noindent where $T_{B,\rm int}$ is the intrinsic brightness temperature. We assume a
typical value for the intrinsic brightness temperature of 10$^{11}$ K \citep[see e.g.][]{lahteenmaki99}, close to the ``equipartition'' value of
5$\times$10$^{10}$ K suggested by \citet{readhead94}. Considering $T_{B,\rm int}$=
10$^{11}$ K in equation~\ref{dopplervar} we obtain $\delta_{var} =$
11 for SBS\,0846$+$513, which is in agreement with the variability Doppler
factor derived for blazars \citep[see e.g.][]{hovatta09}. 
Note that the radio emission used to derive the $\delta_{\rm var}$ and the optical through $\gamma$-ray SED may originate from
different regions of the jet which could have different Doppler factors. 

\subsection{Comparison with $\gamma$-ray blazars}
\label{comparison}

The simultaneous spectrum of SBS\,0846$+$513 collected by Effelsberg during quiescent state in 2011 April showed a flat spectrum  ($\alpha_{\rm r}$
  $\sim$ 0) up to 32 GHz. After the $\gamma$-ray flaring activity occurred in 2012 May and August, significant radio spectral variability was observed (Fig.~\ref{effelsberg}), with the spectrum turning out to be inverted
  ($\alpha_{\rm r}$ $\sim$ $-$0.6). This is a typical blazar-like behaviour,
  which has already been seen in other $\gamma$-ray emitting NLSy1s \citep[see e.g.][]{angelakis13}.
We also found that the SED of SBS\,0846$+$513 in both the flaring and quiescent states is rather typical for a FSRQ, with a Compton
dominance (i.e. the ratio between the luminosity of the high and low-energy peak) of $\sim$5 as well as an X-ray spectrum with a photon index $\Gamma_{\rm X}$ = 1.5--1.6, in agreement with what was pointed out in
\citet{dammando12}. The high-energy part of the source spectrum can be
  modelled by an external Compton component of seed photons from a dust torus,
  similar to FSRQs. Unlike some FSRQs \citep[e.g.~PKS\,0537$-$441;][]{dammando13}, the SEDs of the flaring and quiescent state cannot be
  modelled by changing only the electron distribution; a change of the
  magnetic field strength is also required. Although the fits are not unique, this
  is consistent with the modeling of different activity states of PKS\,0208$-$512, where a change in magnetic filed also seemed to be required
  \citep{chatterjee13}. A significant shift of the synchrotron peak to higher
  frequencies was observed during the 2012 May flaring episode, similar to FSRQs \citep[e.g. PKS\,1510$-$089;][]{dammando11b}.

\begin{figure} 
\centering
\vspace{2.0mm}
\includegraphics[width=7.5cm]{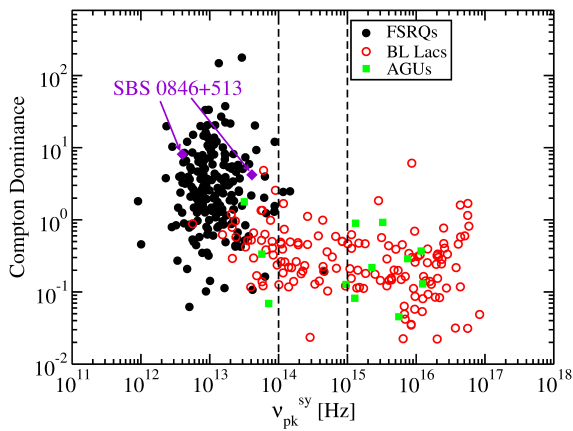}
\caption{Compton dominance vs. peak synchrotron frequency. Filled circles
  represent FSRQs, empty circles represent BL Lac objects, and filled squares
  represent AGN of uncertain type. SBS\,0846$+$513 is plotted as diamonds.}
\label{CD}
\end{figure}

\begin{figure} 
\centering
\vspace{2.0mm}
\includegraphics[width=7.5cm]{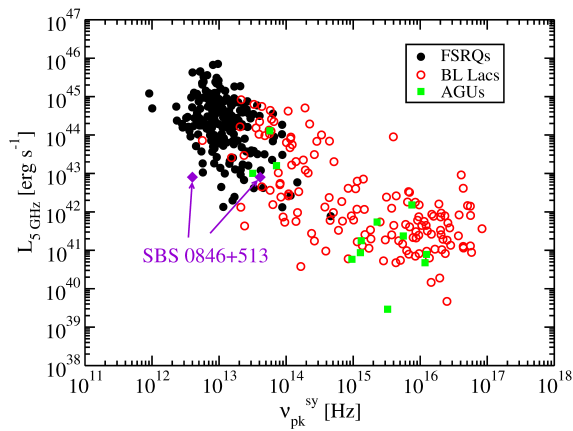}
\caption{Radio luminosity at 5 GHz vs. peak synchrotron luminosity. Filled circles
  represent FSRQs, empty circles represent BL Lac objects, and filled squares
  represent AGN of uncertain type. SBS\,0846$+$513 is plotted as diamonds.}
\label{sequence}
\end{figure}

During the $\gamma$-ray flaring episodes in 2011--2012 SBS\,0846$+$513
reached an apparent isotropic $\gamma$-ray luminosity of 5$\times$10$^{47}$-- 10$^{48}$
erg s$^{-1}$, comparable to those of the bright FSRQs. Recently, a correlation
between Compton dominance ($A_C$) and synchrotron peak frequency in blazars was found
by \citet{finke13}. Considering $\nu_{pk}^{sy}$$\sim$2$\times$10$^{12}$ Hz and
$\sim$2$\times$10$^{13}$ Hz (see Fig.~\ref{0846sed}) for the quiescent
and flaring activity state of SBS\,0846$+$513, respectively, the source seems
to lie in the $A_C$ vs $\nu_{pk}^{sy}$ plot in  
the region occupied by the FSRQs \citep[see Fig.~\ref{CD}; adapted
  from][]{finke13}. This suggests that external Compton scattering is the dominant
mechanism for producing high-energy emission in SBS\,0846$+$513. From the
observation performed by the Effelsberg telescope on 2012 July we derived a 5 GHz radio luminosity $L_{\rm 5 GHz}\sim$8$\times$10$^{42}$ erg s$^{-1}$. Considering the $\nu_{pk}^{sy}$
estimated during the flaring state, SBS\,0846$+$513
occupies the transition region between FSRQs and BL Lac objects in the $L_{\rm
  5 GHz}$ vs $\nu_{pk}^{sy}$ plot, originally proposed by \citet{fossati98} in
the context of the blazar sequence, and then applied by \citet{finke13} to all blazars
detected during the first two years of {\em Fermi} operation (Fig.~\ref{sequence}). It is
interesting to note that considering the $\nu_{pk}^{sy}$ estimated during the
low state, SBS\,0846$+$513 seems to lie slightly outside the blazar
sequence. However, we point out that the determination of the $\nu_{pk}^{sy}$
depends on the amount of data available
across the electromagnetic spectrum and the SED modeling, implying uncertainties on $\nu_{pk}^{sy}$.

The main difference with the powerful FSRQs seems to be the lack of clear
evidence for the accretion disc emission, usually detectable at least during
the low state. However, it is worth mentioning that the accretion disc emission is not visible in all
FSRQs \citep[see e.g.][]{delia03} and $\gamma$-ray emitting NLSy1 \citep[e.g.~PKS\,2004$-$447,][]{abdo09}.

Five radio-loud NLSy1s have been detected by the {\em Fermi}-LAT
\citep{abdo09,dammando12}, stimulating interest in this class of AGN and in
their detection in $\gamma$ rays. Until now however, none of the new candidate
$\gamma$-ray emitting NLSy1s were detected with high significance \citep[see
  e.g.][]{dammando13c,foschini13}. SBS\,0846$+$513 was detected only when it entered a high activity state, having not been detected during the
  first two years of {\em Fermi} operation. This suggests that the flux variability is a key consideration when searching for NLSy1s in $\gamma$
  rays. However, the discoveries of NLSy1s did not always occur during high $\gamma$-ray activity states
\citep[e.g. PKS\,1502$+$036 and PKS\,2004$-$447;][]{abdo09,dammando13b}. Investigating the radio
properties of the first sample of 23 radio-loud NLSy1 presented by \citet{yuan08} we note that the object with the highest radio-loudness, B3\,1044$+$476, has not been detected in $\gamma$ rays by {\em Fermi}-LAT. This indicates that the radio-loudness could be a useful proxy for the jet
production efficiency, but not necessarily for selecting the best candidates
for $\gamma$-ray detection with LAT. In the same way, an apparent
  brightness temperature of $\sim$ 10$^{13}$ K, comparable to that of SBS\,0846$+$513, was observed for TXS\,1546$+$353. Such a high brightness
  temperature could be an indication of Doppler boosted emission from a
  relativistic jet orientated close to our line-of-sight, but from which no
  $\gamma$-ray emission has been detected to date. 
  
\noindent The multi-frequency observations presented here give new clues on the
astrophysical mechanisms at work in SBS\,0846$+$513, though leaving open
questions on the nature of this source. Further multi-wavelength monitoring campaigns are
needed to achieve a complete understanding of the physical processes
occurring in this source and the others detected in $\gamma$ rays by {\em Fermi}-LAT.

\section*{Acknowledgments}

The {\em Fermi} LAT Collaboration acknowledges generous ongoing
support from a number of agencies and institutes that have supported
both the development and the operation of the LAT as well as
scientific data analysis.  These include the National Aeronautics and
Space Administration and the Department of Energy in the United
States, the Commissariat \`a l'Energie Atomique and the Centre
National de la Recherche Scientifique / Institut National de Physique
Nucl\'eaire et de Physique des Particules in France, the Agenzia
Spaziale Italiana and the Istituto Nazionale di Fisica Nucleare in
Italy, the Ministry of Education, Culture, Sports, Science and
Technology (MEXT), High Energy Accelerator Research Organization (KEK)
and Japan Aerospace Exploration Agency (JAXA) in Japan, and the
K.~A.~Wallenberg Foundation, the Swedish Research Council and the 
Swedish National Space Board in Sweden. Additional support for science
analysis during the operations phase is gratefully acknowledged from
the Istituto Nazionale di Astrofisica in Italy and the Centre National
d'\'Etudes Spatiales in France.

We thank the Swift team for making these observations possible, the
duty scientists, and science planners. The OVRO 40-m monitoring program
is supported in part by NASA grants NNX08AW31G and NNX11A043G, and NSF grants AST-0808050 
and AST-1109911. This paper is partly based on observations with the
100-m telescope of the MPIfR (Max-Planck-Institut f\"ur
Radioastronomie) at Effelsberg and the Medicina telescope operated by
INAF--Istituto di Radioastronomia. We acknowledge A. Orlati, S. Righini, and
the Enhanced Single-dish Control System (ESCS) Development Team. FD, MO, MG,
CR acknowledge financial contribution from grant PRIN-INAF-2011. This research has made use of
data from the MOJAVE database that is maintained by the MOJAVE team (Lister et al. 2009, AJ, 137, 3718). IM and VK were supported for this
research through a stipend from the International Max Planck Research School (IMPRS) for Astronomy and Astrophysics at the University of Bonn and
Cologne. We thank D. Horan, P. Bruel, and S. Digel for helpful comments and suggestions.

\end{document}